\documentclass[aps,prl,twocolumn,showpacs,superscriptaddress]{revtex4}

\usepackage[dvips]{graphicx}
\usepackage{amsmath}

\newcommand{\bnmr}{$\beta$-NMR}
\newcommand{\bnqr}{$\beta$-NQR}
\newcommand{\Li}{$^8$Li$^{+}$}

\begin{document}

\title{$\beta$-NMR
of Isolated $^{8}$Li$^{+}$ Implanted into a Thin Copper Film}

\date{\today}

\author{Z.~Salman}
\altaffiliation[Current address: ]{Clarendon Laboratory, Department of
  Physics, Oxford University, Parks Road, Oxford OX1 3PU, UK}
\affiliation{TRIUMF, 4004 Wesbrook Mall, Vancouver, BC, V6T 2A3, Canada}

\author{A.I~Mansour}
\affiliation{Department of Physics, University of Alberta, Edmonton, Canada T6G 2G7}

\author{K.H. Chow}
\email{kimchow@phys.ualberta.ca}
\affiliation{Department of Physics, University of Alberta, Edmonton, Canada T6G 2G7}

\author{M.~Beaudoin} 
\affiliation{Advanced Materials and Process Engineering Lab, 
UBC, Vancouver, BC, V6T 1Z4, Canada}

\author{I. Fan}
\affiliation{Department of Physics, University of Alberta, Edmonton, Canada T6G 2G7}

\author{J. Jung}
\affiliation{Department of Physics, University of Alberta, Edmonton, Canada T6G 2G7}

\author{T.A.~Keeler}
\affiliation{Department of Physics and Astronomy, University of British 
Columbia, Vancouver, BC, V6T 1Z1, Canada}

\author{R.F.~Kiefl}
\affiliation{TRIUMF, 4004 Wesbrook Mall, Vancouver, BC, V6T 2A3, Canada}
\affiliation{Department of Physics and Astronomy, University of British 
Columbia, Vancouver, BC, V6T 1Z1, Canada}
\affiliation{Canadian Institute for Advanced Research}

\author{C.D.P.~Levy}
\affiliation{TRIUMF, 4004 Wesbrook Mall, Vancouver, BC, V6T 2A3, Canada}

\author{R.C. Ma}
\affiliation{Department of Physics, University of Alberta, Edmonton, Canada T6G 2G7}

\author{G.D.~Morris}
\affiliation{TRIUMF, 4004 Wesbrook Mall, Vancouver, BC, V6T 2A3, Canada}

\author{T.J.~Parolin}
\affiliation{Department of Chemistry, University of British Columbia, Vancouver, 
BC, V6T 1Z3, Canada}

\author{D.~Wang}
\affiliation{Department of Physics and Astronomy, University of British 
Columbia, Vancouver, BC, V6T 1Z1, Canada}

\author{W.A.~MacFarlane} 
\affiliation{Department of Chemistry,
University of British Columbia, Vancouver, BC, V6T 1Z3, Canada}

\begin{abstract}

Depth-controlled \bnmr\ was used to study highly spin-polarized \Li\
in a Cu film of thickness 100 nm deposited onto a MgO substrate. The
positive Knight Shifts and spin relaxation data show that \Li\
occupies two sites at low temperatures, assigned to be the
substitutional ($S$) and octahedral ($O$) interstitial sites. Between
50 to 100 K, there is a site change from $O$ to $S$. The temperature
dependence of the Knight shifts and spin-lattice relaxation rates at
high temperatures, i.e. when all the Li are in the $S$ site, is
consistent with the Korringa Law for a simple metal.

\end{abstract}


\maketitle

Nuclear magnetic resonance (NMR) is a powerful tool for investigating
the microscopic magnetic behavior in solid state systems. However,
conventional NMR is often not sensitive enough to investigate thin
film structures and generally cannot be used to study thick conducting
samples. Recently, a high field, beta-detected \Li\ nuclear magnetic
resonance (\bnmr) spectrometer with depth control was developed at
TRIUMF in Vancouver, Canada. This novel instrument utilizes \Li\
(spin 2, lifetime 1.21 s) as the radioactive nuclear spin probe and
provides enough sensitivity to allow NMR studies of thin metal
structures to be carried out\cite{morris04}. These measurements are
aimed at better characterizing the behavior of \Li\ as a prototypical
impurity in metals and to enable investigations of finite-size effects
in these materials. Furthermore, they establish the basis for future
studies of other systems consisting of metal layers such as magnetic
multilayers \cite{Keeler06PB}.

Using the TRIUMF \bnmr\ spectrometer, we recently carried out a
detailed high field, depth resolved, \Li\ \bnmr\ study of a $50$ nm Ag
film\cite{morris04}. Two \bnmr\ resonances were observed at low
temperatures which had Knight Shifts of +120(12) ppm and +212(15) ppm,
implying that \Li\ sits in two different high symmetry sites in the
FCC lattice. Although these shifts are small, the ability to apply
high magnetic fields allowed the two signals to be easily resolved at
all temperatures. Above $\approx 100$ K, the \Li\ makes a transition
from the 212 ppm site to the 120 ppm site, suggesting that the the 212
ppm signal is due to \Li\ residing in the octahedral ($O$)
 site and the 120 ppm
signal is due to \Li\ in the substitutional ($S$) site. The site assignments were
based on comparing the temperature dependence of the $S$ and $O$
signals with other $\beta-$NMR\cite{ackermann83} and
channeling\cite{hoffsass91} experiments in similar systems. For
example, for $^{12}$B in Cu, \bnmr\ cross-relaxation measurements
clearly show that boron occupies the $O$ site at low temperatures, but
moves to the $S$ site at high temperatures \cite{ittermann96}. The
temperature dependences of the Knight Shifts and spin-lattice
relaxation rates (i.e. $1/T_{1}$) were consistent with the Korringa
Law, implying that the \Li\ senses a free-electron like local
electronic susceptibility in the thin Ag film. 

Will similar behavior
be observed for \Li\ implanted into thin films of other ``simple'' FCC
elemental metals such as Cu? In this paper, we experimentally
investigate if such expectations are valid by carrying out \bnmr\
studies of isolated \Li\ in a thin Cu film. In addition, since the
host atoms in Cu have nuclear spin 3/2 and a relatively large
quadrupole moment, this study lays down groundwork for the development
of general techniques that can provide information on the local
structure of the \Li\ site in thin films and near interfaces, such as
``cross-relaxation'' \cite{ittermann96,fullgrabe01,chow_papers}. We
find that similar to Ag\cite{morris04} and (a preliminary study in)
Au\cite{macfarlane03}, \Li\ occupies two sites in Cu at low
temperatures with Knight Shifts of $+120(3)$ and $+182(3)$ ppm,
attributed to the $S$ and $O$ sites respectively.
A transition from the 182 ppm site to the 120 ppm site occurs between 50
and 150 K. The spin lattice relaxation rate at high temperatures
follows a Korringa law as expected for a simple metal.
These measurements will provide a useful reference for future
\bnmr\ and \bnqr\ experiments on samples that use Cu as a thin capping
layer, a substrate, or as part of a multilayered structure.

In the NMR of metals, an important experimental quantity is the
relative shift $\delta$ of the resonance frequency $\nu$ with respect
to the Larmor frequency $\nu_{0}$ of the nucleus in an external
magnetic field $H_{0}$, i.e. $\delta = (\nu - \nu_{0})/\nu_{0}$. In
many ``simple'' metals, $\delta$ is the Knight shift ($K$); it is
independent of temperature and is a consequence of the Fermi contact
interaction of the nuclear spin with the weak Pauli spin paramagnetism
of the conduction electrons\cite{slichter_book}. A second important
experimental quantity is the spin-lattice relaxation rate $1/T_{1}$ of
the nucleus. Under the aforementioned conditions where the contact
hyperfine interaction dominates, the random spin-flip scattering of
the conduction electrons from the nucleus leads to a linear
temperature dependence of the $1/T_{1}$ rate, and hence a product
$T_{1}T$ that is independent of temperature. Furthermore, the
so-called Korringa Law is satisfied: ($T_1 T K^2)/X = 1 $ where $X =
(\gamma_{e}/\gamma_{n})^{2} (h/8\pi^{2}k_{B})$, with $\gamma_{e}$ and
$\gamma_{n}$ denoting the gyromagnetic ratios of the electron and
nucleus respectively. In the case of \Li, the nuclear probe that is
of relevance in this paper, $\gamma_{n}=^{8}\gamma = 6.3015$ MHz/T,
and hence $X = 1.2022 \times 10^{-5}$ s$\cdot$K.

Our current \bnmr\ studies are on a Cu film of thickness $100$ nm
grown via thermal evaporation at a rate of 0.7 nm/s from a 99.999\%
purity Cu source in a pressure of $10^{-7}$ Torr onto a MgO substrate.
A low energy (30.6 keV) beam of highly polarized \Li\ is produced at
the Isotope Separator and Accelerator (ISAC) facility at TRIUMF and
implanted into the sample\cite{morris04,kiefl03,salman04_06}. A large
nuclear polarization ($\approx$70\%) of the \Li\ is generated
in-flight using a collinear optical pumping method. The \bnmr\
spectrometer resides on a high voltage platform which allows the
implantation energy of the \Li\ to be varied between $1-30$~keV when a
suitable positive bias is applied to the platform. These implantation
energies correspond to an average depth from 3 to 100 nm, as
calculated using TRIM.SP\cite{eckstein}. In \bnmr\ the nuclear
polarization, i.e. the quantity of interest, is monitored by
detecting the $\beta$s that are emitted preferentially opposite to the
direction of the \Li\ polarization at the time of decay (i.e. parity
violation). The emitted betas are detected using plastic
scintillation counters forward and backward to the initial \Li\
polarization. The experimentally observed asymmetry $A(t)$ of the
$\beta$ countrates is proportional to the \Li\ nuclear spin
polarization. In our experiment, we are interested in measuring the
magnetic resonance signal and the spin-lattice relaxation rate
$1/T_{1}$: (1) The magnetic resonance is detected by monitoring the
{\it time-averaged} nuclear polarization as a function of the
frequency $\nu$ of a small perpendicular radiofrequency (RF) magnetic
field ($\sim 10^{-4}$ T). The resonance condition is satisfied when
$\nu$ is equal to the Larmor frequency of \Li\ in the internal field.
On resonance, there is a reduction in the nuclear spin polarization;
hence, $A(t)$ and consequentially $A(\nu) = \langle A(t) \rangle$
decreases. (2) The spin-lattice relaxation rate $1/T_1$ is measured
with the RF off. The beam is admitted for 0.5 s every 15 s, and the
evolution of the asymmetry $A(t)$ is measured. Recall that $1/T_1$ is
the rate at which the asymmetry relaxes from its initial value at
$t=0$ to its equilibrium value (i.e. zero polarization).

\begin{figure}
\includegraphics[width=0.7\linewidth]{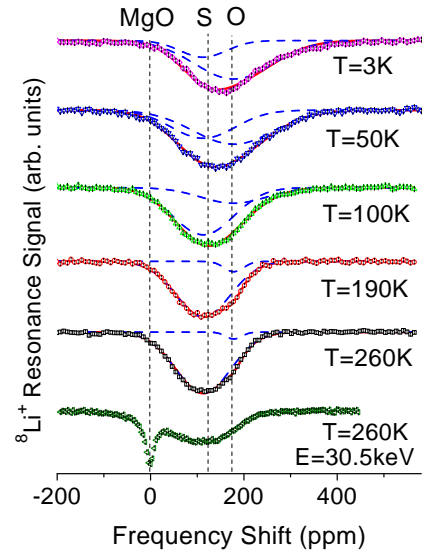}
\caption{ (color online) Representative resonance signals of \Li\ in
the Cu/MgO sample in an applied field of 4.1 T. Each spectrum was
obtained at an implantation energy of $10.6$ keV, except the
bottommost one, which was obtained at the full implantation energy of
30.6 keV. The three vertical dashed lines indicate the peak
frequencies of the MgO, $S$ and $O$ signals. The zero shift in
frequency is taken to be that in MgO at room temperature. The
long-dashed lines indicate the $S$ and $O$ contributions of the signal
at each temperature.}
\label{fig_raw_knight}
\end{figure}

First, we discuss the resonance data. By implanting at the full energy
($30.6$ keV), a significant fraction of the \Li\ stops in the cubic
insulator MgO, providing an {\it in-situ} reference of its Larmor
frequency in the applied field $\nu_{MgO}$ (see
Fig.~\ref{fig_raw_knight}). This is true because the Knight shift of
the \Li\ in an insulator should be zero, and we also expect the
chemical shift to be small\cite{LiChemShifts}. In
Fig.~\ref{fig_raw_knight}, the zero frequency shift corresponds to
$\nu_{MgO}=25.83994(5)$ MHz. Then, at each temperature, the
implantation energy is reduced to $10.6$ keV, a value chosen so that
all of the \Li\ stops in the Cu film; representative spectra at a
number of temperatures in an applied field of 4.1 T are shown in
Fig.~\ref{fig_raw_knight}. The following qualitative features are
apparent: At all temperatures, resonances are observed that are
positively shifted from zero, demonstrating that we are able to detect
the Knight Shifts due to \Li\ in the thin Cu film. At low
temperatures, the lineshape is noticeably asymmetric. In order to
better understand the origin of this asymmetry, we examined the shapes
of the signal with the \Li\ forward and backward polarized, i.e. with
\Li\ initially (primarily) in the $m=+2$ and $m=-2$ state (in our
instrument, this can be done by changing the sense of rotation of the
circularly polarized pumping light). We find that the lineshape is
skewed towards higher frequencies in both situations. This
establishes that the asymmetric lineshape is {\it not} due to
unresolved quadrupole effects since the existence of such interactions
would result in a skewness towards high frequencies in one instance
and a skewness towards lower frequencies in the other
instance\cite{macfarlane03_quadrupole}. Hence, these measurements
rule out a model where the resonance is due to \Li\ stopping in a
single non-cubic site since such a center should experience a
significant quadrupole interaction. Therefore, the asymmetric line at
low temperatures indicates that \Li\ occupies two inequivalent cubic
sites, and each site is characterized by a different positive Knight
shift. However, the shifts are not large enough to be clearly
resolved. As the temperature is raised, the line becomes more
symmetric and the peak frequency shifts to lower values. This
indicates that a change in site has occured.

\begin{figure}
\includegraphics[width=0.65\linewidth]{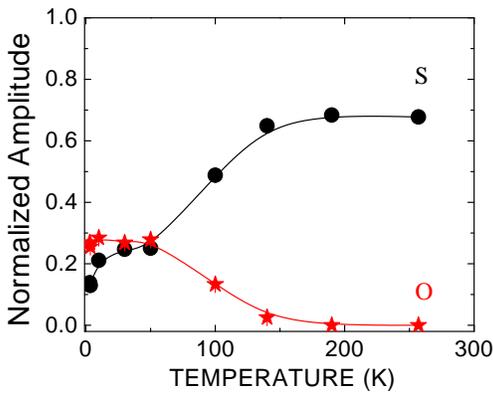}
\caption{ (color online) Temperature dependences of the normalized
amplitudes for an applied field of 4.1 T.}
\label{fig_resonance_amplitude}
\end{figure}
The resonance signals at all temperatures can be fit to a sum of two
Gaussians:
\begin{eqnarray}
A(\nu) = A_{b} &+& 
\frac{A_{S}}{W_{S} \sqrt{\frac{\pi}{2}}}
 \exp\left[-2 \frac{(\nu - \nu_{S})^{2}}{W_{S}^{2}}\right] \nonumber \\
 &+& \frac{A_{O}}{W_{O} \sqrt{\frac{\pi}{2}}}
 \exp\left[-2 \frac{(\nu - \nu_{O})^{2}}{W_{O}^{2}}\right],
\end{eqnarray} 
where $W_{S}$ and $W_{O}$ are the widths and $\nu_{S}$ and $\nu_{O}$
are the peak frequencies. These four parameters are assumed to be the
same at all temperatures. The remaining parameters $A_{b}$ (baseline
term), $A_{S}$, and $A_{O}$ were allowed to vary with temperature. In
anticipation of our assignment of the sites, discussed below, the
subscripts $S$ and $O$ are used to denote the substitutional and
octahedral sites respectively. The fitted widths are $W_{S} = 3.205
\pm 0.006$ kHz and $W_{O} = 4.33 \pm 0.02$ kHz (the error estimates
are entirely statistical\cite{correlation}). 
The Knight Shifts\cite{contraction} can be
obtained from the fitted values of $\nu_{S}$ and $\nu_{O}$, as well as
$\nu_{MgO}$, to be $K_{S} = +120 \pm 3$ ppm and $K_{O} = +182 \pm 3$
ppm. These values are indicated as vertical dashed lines in
Fig.~\ref{fig_raw_knight}. The normalized amplitudes are shown in
Fig.~\ref{fig_resonance_amplitude}, demonstrating that, as the
temperature is increased, there is a thermally activated transition
from the $O$ to the $S$ site. 
Note that the amplitudes do not add up
to unity. 
Possible explanations include : (i) insufficient RF power
to saturate the $S$ and $O$ lines and (ii) the existence of very broad lines 
due to \Li\ stopping in sites
of non-cubic symmetry, and would hence have significant quadrupolar
interaction.

Our results for \Li\ in Cu are similar to that obtained in the Ag film
studied recently\cite{morris04}. For example, the measured low
temperature Knight Shifts in Cu are similar to those in the Ag film
\cite{morris04} of $120$ ppm and $212$ ppm. There, the sites were
attributed to \Li\ located in the substitutional ($S$) and octahedral
($O$) sites respectively. As discussed in the introduction,
such an assignment was made by comparing the data with the ``typical''
temperature dependence seen for light radioactive impurities in
metals, and for $^{12}$B in Cu in particular\cite{ittermann96}. By
analogy, in Cu, we make an assignment of the $120$ ppm signal to the
$S$ site and the $182$ ppm signal to the $O$ site. 
As Fig.~\ref{fig_resonance_amplitude} indicates, \Li\ makes a transition
from the 182 ppm site to the 120 ppm site between $\approx 50$ to 100
K. This is likely a consequence of a thermally activated transition
of the interstitial \Li\ to a nearby vacancy created during the
implantation process. 
It is worthwhile pointing out that Ohsumi {\it et al.} \cite{ohsumi99} also
assigned their high temperature site to the $S$ site after considering 
the \bnmr\ linewidths of \Li\ in a
single crystal of Cu\cite{ohsumi99}. 
However, some of their conclusions are different from ours. 
They propose that $\approx$ 30\% of the \Li\ stops in an $O$
site. In addition, they were not able to detect any changes from 11 K
to 300 K that could be attributed to a site change. The reasons for
the discrepancies with our observations are not clear, but could
partly be due to the fact that their studies were conducted at
significantly lower applied field, and hence are even less able to
resolve two closely spaced lines.
It would require the development of powerful spectroscopic techniques
such as cross-relaxation to unambiguously establish the location of 
\Li\ at all temperatures. 

The linewidths of the \Li\ resonance signals in Cu are significantly
larger than those in Ag\cite{morris04}. The dominant broadening
mechanism of the \bnmr\ signals is from the dipolar broadening
by the host spins\cite{vanvleck48}. 
Hence, the larger linewidths in Cu
compared to Ag are expected since the nuclear moments of the Cu host
atoms are about an order of magnitude greater than those of
Ag\cite{isotope}. Furthermore, the lattice constant of Cu (0.361 nm)
is smaller than that of Ag ($0.408$ nm). The fitted linewidths to
Eq.~1 imply that the second moments of the lines are 3.08 kHz$^{2}$
and 5.63 kHz$^{2}$ for the $S$ and the $O$ sites respectively. The
former is comparable to the prediction of $1.44$ kHz$^{2}$ for \Li\ in
an undistorted $S$ site\cite{ohsumi99}. The extra broadening may be
partly due to the higher RF power used\cite{abragam} here compared to
Ref.~\onlinecite{ohsumi99} ($\sim 3$ times higher).

\begin{figure}
\includegraphics[width=0.80\linewidth]{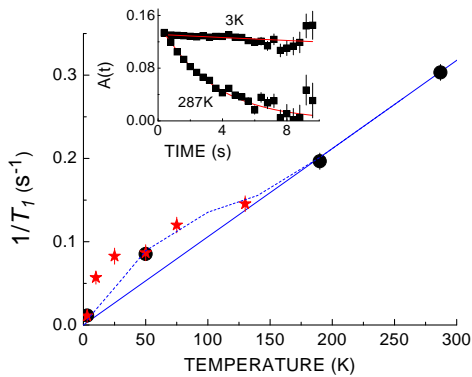}
\caption{ (color online) The temperature dependence of $1/T_{1}$ at
 4.1 T (black circles) and 6.55 T (red stars). The straight line is a
 best fit to the data above 100 K through the origin while
the dashed line is the calculated effective relaxation rate (see text).}
\label{fig_t1}
\end{figure}

We now discuss the spin relaxation data of \Li\ in the Cu film. The
inset in Fig.~\ref{fig_t1} shows examples of the asymmetry $A(t)$; it
is phenomenologically well-described at all temperatures by a single
exponential relaxation function with decay rate $1/T_{1}$. The
temperature dependence of the $1/T_{1}$ rates are shown in
Fig.~\ref{fig_t1}, and are obtained in applied magnetic fields of 4.1
T and 6.5 T. The $1/T_{1}$ values are the same at these two fields, as
expected if the dominant relaxation mechanism at these fields is
Korringa relaxation. The resonance data described above have shown
that at high temperatures (above $\approx$ 100 K), all the \Li\ are in
the $S$ site; here, the $1/T_{1}$ rates are due to the $S$ site only.
They are linear with temperature, and are described by $(T_{1}T)^{-1}
= (1.059 \pm 0.026) \times 10^{-3}$ s$^{-1}\cdot$K$^{-1}$. By
combining this value with the experimentally determined $K_{S}$ of
$+120$ ppm we obtain a Korringa ratio of $T_1 T K^2/X = 1.13 \pm
0.05$, close to the value of unity expected from the
Korringa law. This agreement implies that \Li\ senses a local
susceptibility that is free-electron like. Note that at low
temperatures, there is significant occupation of both the $S$ and $O$
sites and the effective $1/T_{1}$ rates consist of contributions from
\Li\ in both these locations. 
For comparison, the dashed line in 
Fig.~\ref{fig_t1} shows the calculated
effective rates obtained by using the resonance amplitudes from 
Fig.~\ref{fig_resonance_amplitude}, the relaxation rates of the $S$ site
at high temperatures, and the Knight shifts of the $S$ and $O$ sites. 
The calculations are in reasonable agreement with the 
measured results. The deviation at low temperature is most probably due 
to an
additional site with a broad line that was not observed in our
resonance measurements, but contributes to $A(t)$.


By comparison, Ohsumi {\it et al.}\cite{ohsumi99} have reported the
$(T_{1}T)^{-1}$ values for \Li\ in Cu at 20 K, 100 K, and 280 K. They
found that this quantity is nearly independent of temperature with an
average value of (2.4 $\pm$ 0.1) $\times 10^{-3}$
s$^{-1}\cdot$K$^{-1}$. They did not report a value for the Korringa
ratio since measurements of the Knight Shifts were not carried out.
Our value of the Korringa ratio of $1.13 \pm 0.05$ 
is somewhat smaller than
that of $^{63,65}$Cu in Cu of 1.9 \cite{carter77}.

We briefly point out that our preliminary studies in a Cu crystal have
shown that $1/T_{1}$ is independent of the magnetic field above
$\approx 0.07$ T, but is highly field dependent below this value. For
example, at 200 K, the phenomelogical $1/T_{1}$ rates change from the
constant value at $\approx 0.2$ s$^{-1}$ at higher fields to $\approx
0.7$ s$^{-1}$ near zero field. We believe that in this regime, the
dipole-dipole interactions between the \Li\ and the Cu nuclei
contribute significantly to the $1/T_{1}$ rates. The qualitative
explanation for these effects are discussed in numerous references,
including Ref.~\onlinecite{slichter_book,anderson59,hayano79}. We
defer the quantitative discussion of the low field behavior of the
\Li\ relaxation in Cu, as well as other simple metals such as Ag, Au
and Al, to a future publication.

We thank R. Abasalti, B. Hitti, S.R. Kreitzman, and D. Arseneau for
technical assistance and NSERC for support. A. MacDonald collected
and fitted some data.


\begin{thebibliography}{}

\bibitem{morris04}
G.D. Morris {\it et al.}, 
Phys. Rev. Lett. {\bf 93}, 157601 (2004).

\bibitem{Keeler06PB}
T.~A. Keeler {\it et al.},
Physica B {\bf 374-375C}, 79 (2006).

\bibitem{ackermann83}
H. Ackermann {\it et al.} in {\it Hyperfine Interactions of Radioactive
Nuclei} edited by J. Christiansen, Topics in Current Physics Vol. 31
(Springer, Berlin, 1983), p. 291.

\bibitem{hoffsass91}
H. Hoffsass and G. Lindner, Phys. Rep. {\bf 201}, 121 (1991).

\bibitem{ittermann96}
B. Ittermann {\it et al.}, Phys. Rev. Lett. {\bf 77}, 4784 (1996).

\bibitem{fullgrabe01}
M. F\"{u}llgrabe {\it et al.}, 
Phys. Rev. B {\bf 64}, 224302 (2001), 
and references therein.

\bibitem{chow_papers}
K.H. Chow, B. Hitti, R.F. Kiefl, in
 {\it Identification of Defects in Semiconductors},
edited by M. Stavola,
Semiconductors and Semimetals 
vol. 51A
(Academic Press, New York, 1998), p. 137;
B.E. Schultz {\it et al.}, 
Phys. Rev. B {\bf 72}, 33201 (2005);
B.E. Schultz {\it et al.}, 
Phys. Rev. Lett. {\bf 95}, 86404 (2005);
K.H. Chow {\it et al.}, 
Phys. Rev. Lett. {\bf 87}, 216403 (2001);
K.H. Chow {\it et al.}, 
Phys. Rev. B {\bf 51}, 14762 (1995).

\bibitem{macfarlane03}W.A. MacFarlane {\it et al.}, 
Physica B {\bf
326}, 213 (2003).


\bibitem{slichter_book} C. P. Slichter, {\it Principles of Magnetic Resonance}, 3rd 
ed. (Springer, Berlin, 1990).

\bibitem{kiefl03}	
R.F. Kiefl {\it et al.}, 
Physica B {\bf 326}, 189 (2003).

\bibitem{salman04_06}Z. Salman {\it et al.}, 
Phys. Rev. B {\bf 70}, 104404 (2004);
Z. Salman {\it et al.}, 
Phys. Rev. Lett. {\bf 96}, 147601 (2006).

\bibitem{eckstein} 
W.~Eckstein, {\em Computer Simulation of Ion-Solid Interactions} 
(Springer, Berlin, 1991).

\bibitem{LiChemShifts}
e.g. C.P. Grey and N. Dupr\`e, Chem. Rev. {\bf 104},
4493 (2004);
Z. Xu and J.F. Stebbins, Sol. St. Nucl. Mag. Reson. {\bf 5}, 103 (1995).

\bibitem{macfarlane03_quadrupole}
e.g. see W.A. MacFarlane {\it et al.}, 
Physica B {\bf 326}, 209 (2003).

\bibitem{correlation}
The correlation coefficients between the different parameters 
are low, e.g. -0.0042 and -0.32746 between the centers and the widths
of the resonances, respectively.
This indicates the model used is not overparameterized. 

\bibitem{contraction}
The assumption that $\nu_{S}$ and $\nu_{O}$ are independent
of temperature ignores the influence of lattice
contraction on the Knight Shifts. 
From analogy with \Li\ in Ag\cite{morris04}, 
we believe this will be a small effect.

\bibitem{ohsumi99}
F. Ohsumi {\it et al.}, 
Hyp. Interact. {\bf 121}, 419 (1999). 

\bibitem{vanvleck48}
J.H. van Vleck, Phys. Rev. {\bf 74}, 1168 (1948). 

\bibitem{isotope}
Cu has two isotopes with
a nuclear spin $I=3/2$:
$^{63}$Cu (abundance 69.2\%) and $^{65}$Cu (abundance 30.8\%) with
gyromagnetic ratios $11.33$ MHz/T and $12.10$ MHz/T respectively. 
By contrast, Ag has
two isotopes with a nuclear spin $I=1/2$:
$^{107}$Ag (abundance 51.8\%) and
$^{109}$Ag (abundance 48.2\%)
with gyromagnetic ratios $-0.173$ MHz/T and 
$-0.1992$ MHz/T respectively. 

\bibitem{abragam}
A. Abragam, {\it The Principles of Magnetic Resonance}
(Oxford University Press, 1961) p. 45. 

\bibitem{carter77}
G.C. Carter, L.H. Bennett, and D.J. Kahan,
{\it Metallic Shifts in NMR} (Pergamon, Oxford, 1977).


\bibitem{anderson59}
A.G. Anderson and A.G. Redfield, 
Phys. Rev. {\bf 59}, 583 (1959).

\bibitem{hayano79}
R.S. Hayano {\it et al.},
Phys. Rev. B {\bf 20}, 850 (1979). 



\end{thebibliography}
\end{document}